# Formation of optical vortices using coherent laser beam arrays

Li-Gang Wang,[1, 2 *] Li-Qin Wang,[1] and Shi-Yao Zhu [1, 2, 3]

[1] *Department of Physics, Zhejiang University, Hangzhou 310027, China*
[2] *Department of Physics, The Chinese University of Hong Kong, Shatin, N. T., Hong Kong, China*
[3] *Department of Physics, Hong Kong Baptist Univerisity, Kowloon Tong, Hong Kong, China*
[*] *Corresponding author:sxwlg@yahoo.com.cn*

**Abstract:** We present a novel proposal to generate an optical vortex beam by using the coherent-superposition of multi-beams in a radial symmetrical configuration. In terms of the generalized Huygens-Fresnel diffraction integral, we have derived the general propagation expression for the coherent radial laser arrays. Based on the derived formulae, we have analyzed the effects of the beamlet number $N$, the separation distance $\rho$ of the beamlets and the topological charge $m$ on the intensity and phase distributions of the resulted beams. Our simulation results show that optical vortices could be efficiently formed and generated due to the interference and superposition effect of all the beamlets, during the propagation process of the coherent radial laser arrays with the initial well-organized phase distributions through the free space. In the focusing system, the resulted beam near the focusing plane has the strong rotation effect with the phase helicity.

## References and links

1. J. F. Nye and M. V. Berry, "Dislocations in wave trains," Proc. R. Soc. London Ser. A **336**, 165 (1974).
2. J. F. Nye, *Natural focusing and fine structure of light：caustics and wave dislocations (*Institute of Physics Publishing, 1999).
3. A. Mair, A. Vaziri, G. Weihs, and A. Zeilinger, "Entanglement of the orbital angular momentum states of photon," Nature **412**, 313 (2001).
4. L. Allen, M. W. Beijersbergen, R. J. C. Spreeuw, and J. P. Woerdman, "Orbital angular momentum of light and the transformation of Laguerre-Gaussian laser modes," Phys. Rev. A **45**, 8185 (1992).
5. P. Senthilkumaran, "Optical phase singularities in detetion of laser beam collimation," Appl. Opt. **42**, 6314-6320 (2003).
6. K. Dholakia, N. B. Simpson, M. J. Padgett, and L. Allen, "Second-harmonic generation and the orbital angular momentum of light," Phys. Rev. A **54**, R3742 (1996).
7. A. Ashkin, "Forces of a single-beam gradient laser trap on a dielectric sphere in the ray optics regime," Biophys. J. **61**, 569 (1992).
8. K. T. Gahaghan and G. A. Swartzlander Jr., "Optical vortex trapping of particles," Opt. Lett. **21**, 827 (1996).
9. V. Westphal and S. W. Hell, "Nanoscale Resolution in the Focal Plane of an Optical Microscope," Phys. Rev. Lett. **94**, 143903 (2005).
10. M. D. Levenson, T. J. Ebihara, G. Dai, Y. Morikawa, N. Hayashi, and S. M. Tan, "Optical vortex masks for via levels," J. Microlithogr. Microfabr. and Microsyst. **3**, 293-304 (2004).
11. H. H. Arnaut and G. A. Barbosa, "Orbital and intrinsic angular momentum of single photons and entangled pairs of photons generated by parametric down-conversion," Phys. Rev. Lett. **85**, 286 (2000)
12. S. Franke-Arnold, S. M. Barnett, M. J. Padgett, and L. Allen, "Two-photon entanglement of orbital angular momentum states," Phys. Rev. A **65**, 033823 (2002).
13. G. Foo, D. M. Palacios, and G. A. Swartzlander, "Optical vortex coronagraph," Opt. Lett. **30**, 3308 (2005).
14. J. H. Lee, G. Foo, E. G. Johnson, and G. A. Swartzlander, Jr., "Experimental verification of an optical vortex coronagraph," Phys. Rev. Lett. **97**, 053901 (2006).
15. E. Abramochkin, V. Volostnikov, "Beam transformations and nontransformed beams," Opt. Commun. **83**, 123 (1991).

16. M. W. Beijersbergen, L. Allen, H. E. L. O. van der Veen, J. P. Woerd-man, "Astigmatic laser mode converters and transfer of orbital angular momentum," Opt. Commun. **96**, 123 (1993).
17. N. R. Heckenberg, R. McDuff, C. P. Smith, and A. G. White, "Generation of optical phase singularities by computer-generated holograms," Opt. Lett. **17**, 211-233 (1992).
18. M. Beijersbergen, R. P. C. Coerwinker, M. Kristensen, and J. P. Woerdman, "Helical-wavefront laser beams produced with a spiral phaseplate," Opt. Commun. **112**, 321 (1994).
19. C. Rotschild, S. Zommer, Sh. Moed, O. Hershcovitz, and S. G. Lipson, "Adjustable Spiral Phase Plate," Appl. Opt. **43**, 2397 (2004).
20. V. V. Kotlyar, and A. A. Kovalev, "Fraunhofer diffraction of the plane wave by a multilevel (quantized) spiral phase plate," Opt. Lett. **33**, 189 (2008).
21. Y. Izdebskaya, V. Shvedov, and A. Volyar, "Generation of higher-order optical vortices by a dielectric wedge," Opt. Lett. **30**, 2472 (2005).
22. X. C. Yuan, B. P. S. Ahluwalia, H. L. Chen, J. Bu, and J. Lin, "Generation of high-quality optical vortex beams in free-space propagation by microfabricated wedge with spatial filtering technique," Appl. Phys. Lett. **91**, 051103 (2007).
23. K. O'Holleran, M. J. Padgett, and M. R. Dennis, "Topology of optical vortex lines formed by the interference of three, four, and five plane waves," Opt. Express **14**, 3039 (2006).
24. S. Vyas, and P. Senthilkumaran, "Interferometric optical vortex array generator," Appl. Opt. **46**, 2893 (2007).
25. M. S. Petrović, "Counterpropagating mutually incoherent vortex-induced rotating structures in optical photonic lattices," Opt. Express **14**, 9415 (2006).
26. Y. Izdebskaya, T. Fadeyeva, V. Shvedov, and A. Volyar, "Vortex-bearing array of singular beams with very high orbital angular momentum," Opt. Lett.**31**, 2523(2006)
27. K. M. Abramski, A. D. Colley, H. J. Baker, and D. R. Hall, "High-power two-dimensional waveguide CO2 laser arrays," IEEE J. Quantum Electron. **32**, 340 (1996).
28. W. D. Bilida, J. D. Strohschein and H. J. J. Seguin, "High-power 24 channel radial array slab RF-excited carbon dioxide laser," in Gas and Chemical Laser and Applications II, R. C. Sze and E. A. Dorko, eds., Proc. SPIE **2987**, 13 (1997).
29. J. D. Strohschein, H. J. J. Seguin, and C. E. Capjack, "Beam propagation constants for a radial laser array," Appl. Opt. **37**, 1045 (1998).
30. B. Lü and H. Ma, "Beam propagation properties of radial laser arrays," J. Opt. Am. A. **17**, 2005 (2000)
31. V. Eckhouse, A. A. Ishaaya, L. Shimshi, N. Davidson, and A. A. Friesem, "Intracavity. coherent addition of 16 laser distributions," Opt. Lett. **31**,350 (2006)
32. B. Lü and H. Ma, "Beam combination of a radial laser array: Hermite–Gaussian model," Opt. Commun. **178**, 395 (2000).
33. C. L. Zhao, L. G. Wang, X. H. Lu, "Radiation forces on a dielectric sphere produced by highly focused hollow Gaussian beams," Phys. Lett. A. **363**, 502 (2006).
34. A. Ashkin, "Trapping of Atoms by Resonance Radiation Pressure," Phys. Rev. Lett. **40**, 729-732 (1978).
35. S. Chu, J. E. Bjorkholm, A. Ashkin, A. Cable, "Experimental observation of optically trapped atoms," Phys. Rev. Lett. **57**, 314-317 (1986).
36. S. A. Collins, "Lens-System Diffraction Integral Written in Terms of Matrix Optics," J. Opt. Soc. Am. **60**, 1168 (1970).
37. Q. Lin and L. Wang, "Collins formula and tensor ABCD law in spatial-frequency domain," Opt. Commun. **185**, 263 (2000).
38. V. S. Ilchenko, M. Mohageg, A. A. Savchenkov, A. B. Matsko, and L. Maleki, "Efficient generation of truncated Bessel beams using cylindrical waveguides," Optics Express **15**, 5866-5871 (2007).
39. V. A. Pas'ko, M. S. Soskin, and M. V. Vasnetsov, "Transversal optical vortex," Opt. Commun. **198**, 49-56 (2001).

## 1. Introduction

Optical vortices, which contain topological wave front dislocations [1], have attracted intensive attentions in many branches of classical and quantum physics [2-3]. The wave front dislocation refers to a continuous line on which the wave phase is undetermined (singular) and its field amplitude vanishes or, in a mathematical form, they are zeros of complex functions with a nonzero phase circulation around them. Owing to this circulation, the optical wave carries the orbital angular momentum [4]. According to the classification introduced by Nye and Berry [1], there are mainly two types of phase singularities: screw wavefront dislocation and edge dislocation, although the mixed screw-edge dislocation occurs in most situations. Optical vortices have many important applications, e. g., optical testing [5], nonlinear optics

[6], optical tweezers [7-8], high-resolution fluorescence microscopy [9], lithography [10], quantum entanglement [3, 11, 12], and stellar coronagraph [13, 14].

Nowadays, optical vortices can be generated by various methods such as mode conversions [15-16], computer generated holograms [17], spiral phase plates [18-19] and multi-level spiral phase plate [20], and optical wedges [21, 22]. Recently, the interference of several plane waves was used to generate the optical vortices [23]. Vyas and Senthilkumaran [24] have proposed the modified Michelson interferometer and the modified Mach-Zehnder interferometer for producing optical vortex arrays. Petrović [25] have found the stable rotating structures in the optical photonic lattices by using the rotating counterpropagating incoherent self-trapped vortex beams. Izdebskaya et al. [26] have experimentally obtained the singular beams with high orbital angular momentum by using an array of vortex beams.

In this paper, a radial beam array composed of coherent Gaussian beamlets is proposed to generate optical vortices. In the current scheme, we arrange the fundamental Gaussian beams with the initial well-ordered phases in a radial symmetric configuration, and then we consider the propagation of the radial beam array passing through the first-order linear optical systems and found the stable optical vortices formed at the far-field region in the free space and near the focusing plane in the lens optical systems. It should be mentioned that the laser arrays have been investigated widely because of their potential applications in high-energy weapon and atmospheric optical communication [27-31]. In the previous studies [27-32], the beam arrays in the radial or rectangular symmetrical configurations are investigated in in phase-locked and non-phase-locked cases. For the phase-locked case it usually refers to the coherent superposition of all beamlets with the same initial phases, while for the non-phase locked case it refers to the incoherent addition of all beamlet [30]. However, unlike the previous phase-locked radial laser arrays, each beamlet of the coherent radial beam arrays considered here has the different initial phase in a well-ordered distribution. It is found that the optical vortices could be formed during the propagation of such radial symmetric beam arrays. Our results have useful application in the optical manipulation of micro-sized particles [7-8,33] and atomic trapping and guiding [34-35].

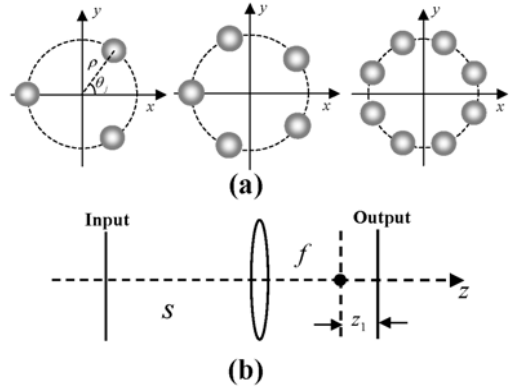

Fig. 1. (a) Schematic of a radial laser arrays with $N=3,5,8$ identical fundamental Gaussian beams, which are uniformly located on a ring with radius $\rho$ at the azimuth angle $\theta_j$ ( $j=1,\cdots,N$ ). (b) Schematic of a focusing lens optical system.

## 2. A radial laser array and its propagation

Consider a radial laser array consisted of $N$ identical fundamental Gaussian beams, which are located symmetrically on a ring with radius $\rho$, as illustrated in Fig.1(a). Each beamlet is assumed to be symmetrical with the waist width $w_0$ and different initial phases $\varphi_j$, and the displacement coordinates $(a_j, b_j)$ of each beamlet's center are $(\rho\cos\theta_j, \rho\sin\theta_j)$, where

$\theta_j = \pi(2j-1)/N$ is the azimuth angle of the $j$th beamlet and the radius $\rho$ controls the separation distance among these beamlets. The initial beam array is located at the input plane of $z = 0$, and propagates through an axial symmetrically optical system. For the $j$th initial Gaussian beamlet, its field distribution in the rectangular coordinates is defined by

$$E_j(x_1, y_1; z=0) = G_0 e^{-\frac{(x_1-a_j)^2+(y_1-b_j)^2}{w_0^2}} e^{i\varphi_j}, \tag{1}$$

where $j = 1, 2, \cdots, N$ denotes the list of the beamlets, the initial phase $\varphi_j = \pi m(2j-1)/N$, $m$ is related to the topological charge of the resulted beam, $w_0$ the waist width of the Gaussian beam, and $G_0$ a constant. The output field distribution of each beamlet passing through a first-order optical system is described by the generalized Huygens-Fresnel diffraction integral [36-37], which can be written as

$$E_j(x_2, y_2; z) = -\frac{ik}{2\pi B} e^{ikL_0} \int_{-\infty}^{\infty}\int_{-\infty}^{\infty} E_j(x_1, y_1; 0) e^{\frac{ik}{2B}[A(x_1^2+y_1^2)-2(x_1x_2+y_1y_2)+D(x_2^2+y_2^2)]} dx_1 dy_1, \tag{2}$$

where $k = 2\pi/\lambda$ is the wave number of the incident Gaussian beamlets with wavelength $\lambda$. The first-order linear optical system is denoted by the ray transfer matrix $\begin{pmatrix} A & B \\ C & D \end{pmatrix}$, and $L_0$ is the axial optical path length from the input plane ($z=0$) to the output plane ($z$). On submitting from Eq. (1) into Eq. (2) and after tedious integral calculation, we can obtain the expression of each beamlet's field distribution on the output plane as follows:

$$E_j(x_2, y_2; z) = -G_0\left(\frac{iZ_r}{\xi}\right) e^{ikL_0} e^{i\varphi_j} e^{\frac{ikA(a_j^2+b_j^2)}{2\xi} - \frac{ik(a_jx_2+b_jy_2)}{\xi}} e^{\frac{ik\eta(x_2^2+y_2^2)}{2}}, \tag{3}$$

where $Z_r = kw_0^2/2$ is the Rayleigh distance, $\xi = B - iAZ_r$, and $\eta = (D - iCZ_r)/\xi$. Equation (3) describes a decentered Gaussian beam which retains the Gaussian property unchanged after propagating through an unapertured *ABCD* system.

Unlike the previous considerations [29-30], here we consider the case that the radial beam array is combined with the identical fundamental Gaussian beams having different initial phases. During the propagation, each beamlet will interfere with each other, which induces the novel properties with the dark-intensity center, the rotation of the resulted beam and helical phase structure. Due to the fact that all the beamlets are coherent superposition, the resulted field distribution of the radial laser array on the output plane could be expressed as

$$E(x_2, y_2; z) = \sum_{j=1}^{N} E_j(x_2, y_2; z)$$

$$= -G_0\left(\frac{iZ_r}{\xi}\right) e^{ikL_0} \exp\left[\frac{ikA\rho^2}{2\xi}\right] \exp\left[\frac{ik\eta(x_2^2+y_2^2)}{2}\right] \sum_{j=1}^{N} \exp\left[-\frac{ik(a_jx_2+b_jy_2)}{\xi} + i\varphi_j\right]$$

$$= -G_0\left(\frac{iZ_r}{\xi}\right) e^{ikL_0} \exp\left[\frac{ikA\rho^2}{2\xi}\right] \exp\left[\frac{ik\eta(x_2^2+y_2^2)}{2}\right]$$

$$\times \sum_{j=1}^{N} \exp\left\{-\frac{ik\rho\{x_2 \cos[\pi(2j-1)/N] + y_2 \sin[\pi(2j-1)/N]\}}{\xi} + i\frac{\pi m}{N}(2j-1)\right\}$$

(4)

Using Eq. (4), the corresponding intensity distribution on the output plane reads as

$$I(x_2, y_2; z) = E(x_2, y_2; z) E^*(x_2, y_2; z), \tag{5}$$

and its phase distribution could be readily obtained from

$$\varphi = \tan^{-1}\{\text{Im}[E(x_2, y_2; z)]/\text{Re}[E(x_2, y_2; z)]\}, \quad \varphi \in [0, 2\pi). \tag{6}$$

Equations (4)-(6) provide a general description of the radial beam arrays passing through an axial symmetrically first-order optical system. In the following discussions, we use Eqs. (4)-(6) to study the propagation properties of such radial beam arrays and how the optical vortex produces from the coherent superposition of the radial beam arrays. The location of a vortex within the intensity distribution could be found by examining the phase structure around each point in the phase distribution, a change from 0 to $2\pi$ around a point indicating a screw-type vortex or a phase discontinuous line indicating an edge dislocation. In the following simulations, we take the beam array's parameters as follows: $\lambda = 632.8$ nm, $w_0 = 1$ mm, and consequently $Z_r \approx 5$ m.

## 3. Numerical results and discussions

Figures 2, 3 and 4 show the typical evolutions of the intensity and phase distributions for the different radial beam arrays propagating through the free spaces. The ray transfer matrix of the free space is given by $\begin{pmatrix} A & B \\ C & D \end{pmatrix} = \begin{pmatrix} 1 & z \\ 0 & 1 \end{pmatrix}$, where $z$ is the propagation distance. From Figs. 2, 3 and 4, we find that each beamlet of the radial beam array is initially separated from each other, and gradually interferes with each other during the propagating process. From the intensity evolutions in Figs. 2-4, it is clear seen that the regions with the maximal intensities gradually rotates. For example, as shown in Fig. 2, the radial beam array is initially composed of three identical separated Gaussian beams with different initial phases; as the propagating distance is increasing, the interference among these three beamlets leads to the rotation phenomena in the intensity distribution of the resulted beam (see the changes of the dashed triangles in Fig. 2). From Figs. 2, 3 and 4, it is shown that the phase evolution of the radial beam array reveals the physical nature of the intensity evolution. Clearly, there is always a screw-type phase singularity at the center position (0,0) around which the phase changes from 0 to $2\pi$. It is easily to find that there are other phase singularities shown in the below of Fig. 2(d), which also correspond to the other zero-intensities. Therefore the intensities of the radial beam array at the center point or other vortex cores are always equal to zero, i. e., the resulted beam becomes a annular hollow beam at the far field region. During the propagating process, for the beam array with small $N$, the initial phase dislocations evolve into the isolated phase singularities [i. e., the so-called screw-type dislocations, see the below parts from Fig. 2(a) to 2(d)], and around each phase singularity the beam intensity is close to zero. For the beam array with much larger $N$, by suitably adjusting the parameter $\rho$ (which controls the separation distance among these beamlets), the resulted beam may form an isolated dark-hollow beam in the inner region, and by a carefully observation one could find a phase singular circle denoted by the dashed circle in the phase distribution of Fig. 4(d), which corresponds to the phase edge-like dislocation for larger $N$; and at the outer part of the resulted beam the phase discontinuities still evolve into the isolated phase singularities, therefore the resulted beam gradually forms an annular beam with a zero-intensity center in the inner part and becomes the shape like a gear wheel at the outer part [see Figs. 4(a) to 4(d)]. Meanwhile, the inner dark-hollow beam is isolated from the other intensity distribution [see Figs. 4(d), 7(b) and 7(c)]. Here it should be pointed out that it is easy to observe the singular phase circle in Figs. (7b) and (7c). Actually, such beams with phase singularities are so-called optical vortices which could have the rotation phenomena and have the orbital angular moments [2, 4]. We emphasize that the first inner dark-hollow beam may contain the most energy of the initial input array and with the screw phase structure, and the intensity distributions of the optical vortices generated by this method, i. e., the radial beam array by the coherent combination of Gaussian beams with different initial phases, become stable in the far field region of the free space [see the parts (c) and (d) of Figs. 2, 3 and 4].

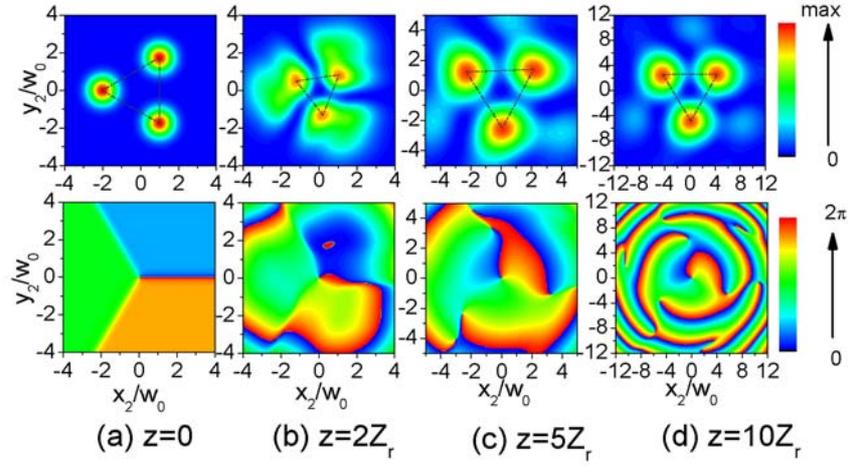

Fig. 2. Evolution of the intensity (upper) and phase (below) distributions of the radial beam array ($N=3$) at different propagating distances: (a) $z=0$, (b) $z=2Z_r$, (c) $z=5Z_r$, and (d) $z=10Z_r$, with other parameters $m=1$ and $\rho=2w_0$.

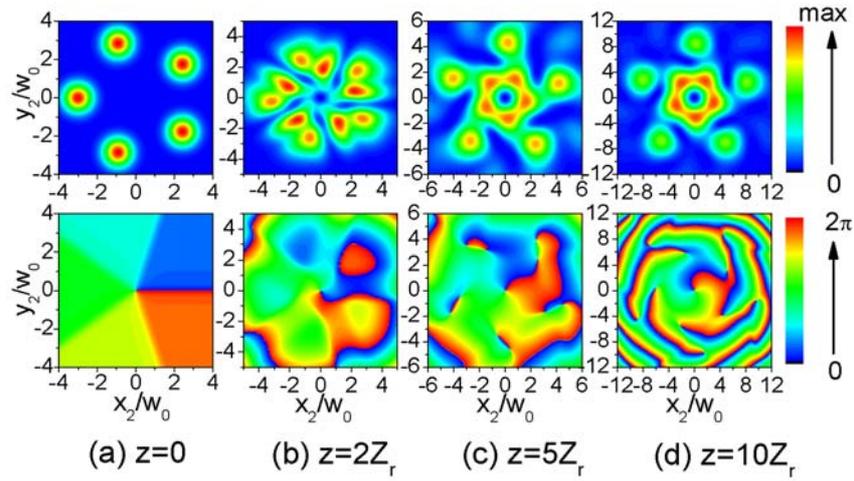

Fig. 3. Evolution of the intensity (upper) and phase (below) distributions of the radial beam array ($N=5$) at different propagating distances: (a) $z=0$, (b) $z=2Z_r$, (c) $z=5Z_r$ and (d) $z=10Z_r$, with other parameters $m=1$ and $\rho=3w_0$.

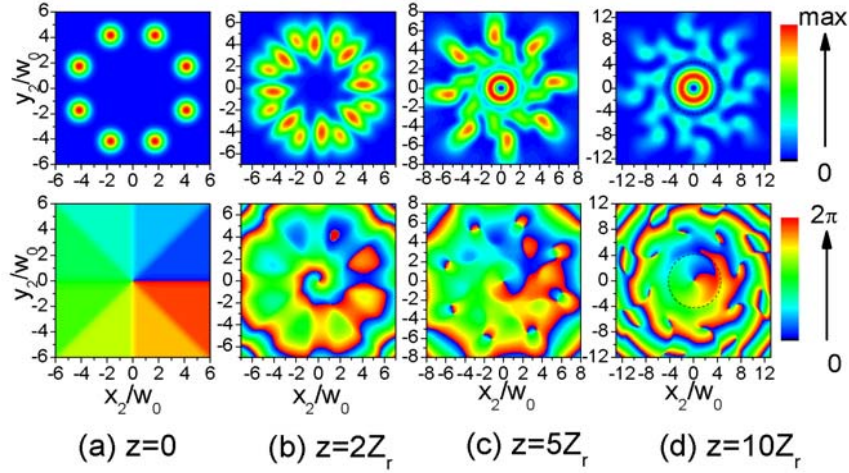

Fig. 4. Evolution of the intensity (upper) and phase (below) distributions of the radial beam array ( $N=8$ ) at different propagating distances: (a) $z=0$, (b) $z=2Z_r$, (c) $z=5Z_r$ and (d) $z=10Z_r$, with other parameters $m=1$ and $\rho=4.5w_0$. In (d), the dashed circle in the phase diagram denotes the edge-like phase dislocation, which corresponds to a zero-intensity circle in the intensity profile.

In order to study the far-field propagation properties of the radial coherent beam arrays, we use a lens transformation system to image the far-field intensity distributions on the focusing plane. As shown in Fig. 1 (b), the ray transfer matrix for an apertureless lens system between the input and the output planes could be given by

$$\begin{bmatrix} A & B \\ C & D \end{bmatrix} = \begin{bmatrix} 1 & z_1+f \\ 0 & 1 \end{bmatrix} \begin{bmatrix} 1 & 0 \\ -1/f & 1 \end{bmatrix} \begin{bmatrix} 1 & s \\ 0 & 1 \end{bmatrix} = \begin{bmatrix} -z_1/f & f+(1-\alpha)z_1 \\ -1/f & 1-\alpha \end{bmatrix}, \quad (7)$$

where $s$ is the distance between the input plane to the thin lens, $f$ is the focus length, and $z_1$ is the distance from the focusing plane to the output plane. Therefore, the far-field properties in the free space can be studied directly by the intensity and phase distributions on the focusing plane ( $z_1=0$ ).

In Fig. 5, we observe a typical intensity rotation process before and after the focusing plane ( $z_1=0$ ). It is clear seen that the resulted beam is clockwise rotating around the propagating axis $z$ and transferring its energy into the inner dark-hollow beam before the focusing plane, while after the focusing plane the resulted beam is still clockwise rotating but transferring its energy from the inner dark-hollow beam to the outside. Near the focusing plane the most energy of the beam array is transferring into the inner dark-hollow vortex beam [see Fig. 5(d)]. Therefore the generated optical vertex by this coherent radial beam array has orbital angular moment and near the focusing plane, the resulted vortex beam may be used to manipulate and rotate the small particles [7-8].

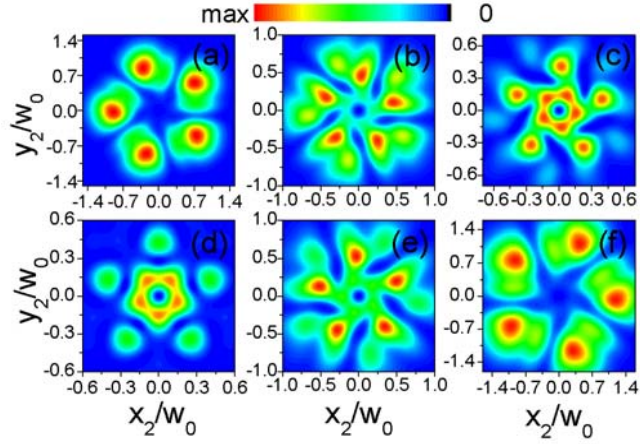

Fig. 5. The output intensity distributions of the radial beam array ($N = 5$) through a lens near the focusing plane: (a) $z_1 = -0.15Z_r$, (b) $z_1 = -0.1Z_r$, (c) $z_1 = -0.05Z_r$ and (d) $z_1 = 0$, (e) $z_1 = 0.1Z_r$ and (f) $z_1 = 0.2Z_r$ with other parameters $m = 1$ and $\rho = 3w_0$, $s = 0.2Z_r$ and $f = 0.5Z_r$.

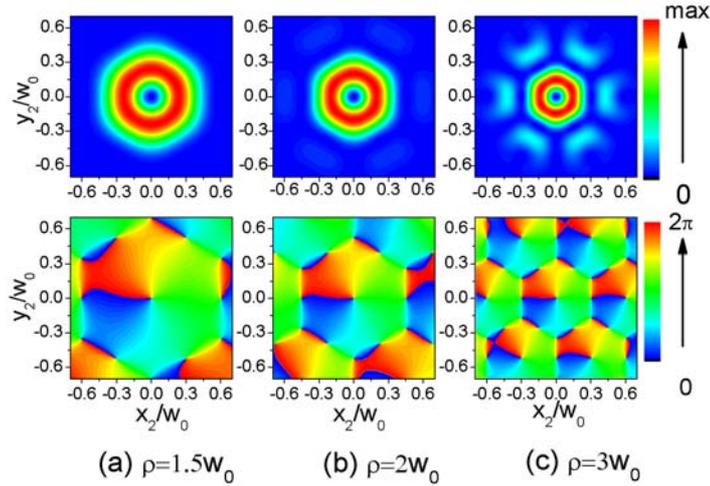

Fig. 6. The intensity (upper) and phase (below) distributions of the different radial beam arrays with different parameters (a) $\rho = 1.5w_0$, (b) $\rho = 2w_0$ and (c) $\rho = 3w_0$ at the focusing plane $z_1 = 0$, with other parameters $N = 6$, $m = 1$, $s = 0.2Z_r$ and $f = 0.5Z_r$.

In order to obtain the high-quality optical vertex with the dark-intensity center on the focusing plane, one has to adjust the parameter $\rho$ which controls the separation distance among these beamlets. As an example shown in Fig. 6, it is found that in this case when we choose a suitable value of the parameter $\rho = 2w_0$, the most light energy is transferred into the inner annular dark-hollow beam with few sidelobes; if the parameter $\rho$ becomes smaller, although there is no sidelebe in the focusing intensity distribution, the inner annular dark-hollow beam has a much large scale; if the parameter $\rho$ becomes larger, although the scale of the inner annular dark-hollow beam has a much smaller size, more and more light energy is transferred into the sidelobes of the intensity profile. Similar effects are existed for the radial beam arrays with different $N$.

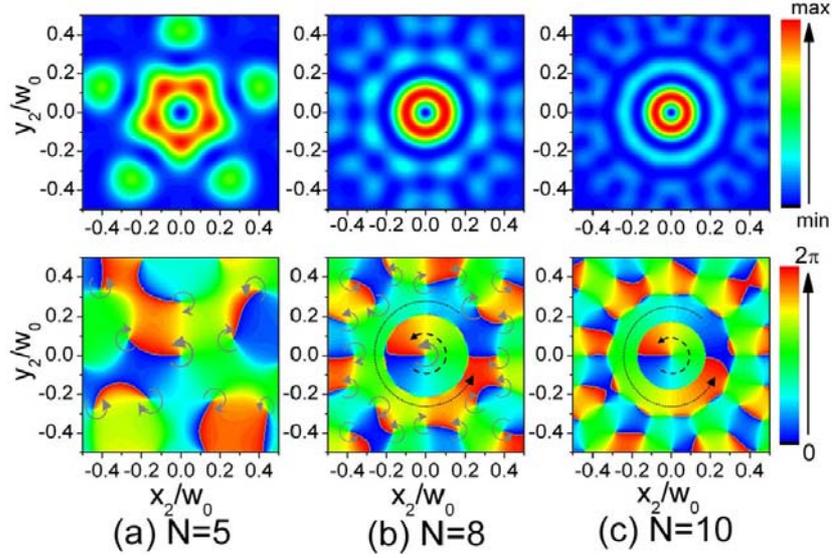

FIG. 7. The intensity (upper) and phase (below) distributions for the different radial beam arrays with (a) $N = 5$ and $\rho = 3w_0$, (b) $N = 8$ and $\rho = 4.5w_0$, and (c) $N = 10$ and $\rho = 5.5w_0$, at the focusing plane $z_1 = 0$. Note that in (a) and (b) the curved gray arrows encircle the phase singularities that correspond to the dark intensity regions (upper). The other parameters are $s = 0.2Z_r$, $f = 0.5Z_r$ and $m = 1$.

Figure 7 shows the intensity and phase distributions on the focusing plane for different coherent radial beam arrays. In Fig. 7 (a) and (b), the gray curved arrows in the phase distributions denote the screw phases around the phase singularities, and each of these phase singularities corresponds to the zero-intensity distribution and a vortex core. With the increasing of the number $N$ of the Gaussian beamlets and the suitable choice of the parameter $\rho$, one may find double-annular or multi-annular dark-hollow beam in the inner part of the intensity distribution, i. e., the center vortex core forms the innerest dark-hollow vortex beam and the second circular vortex strip forms the second dark-hollow vortex beam, and so on [see Fig. 7 (b) and 7(c) and 8(a)]. The dark-intensity pattern between two annular bright patterns always corresponds to a circular singular-phase dislocation. For example, in Figs. 7 (b) and 7(c), the inner dash-curved arrows correspond to the first inner annular dark-hollow vortex beam and the second dot-curved arrows form the second annular vortex pattern; and between these two annular vortex patterns there is a circle of the edge-type phase dislocation corresponding to the circular dark pattern. Therefore it is also possible to obtain the multi-annular hollow vortex beams with several homocentric phase helical structures.

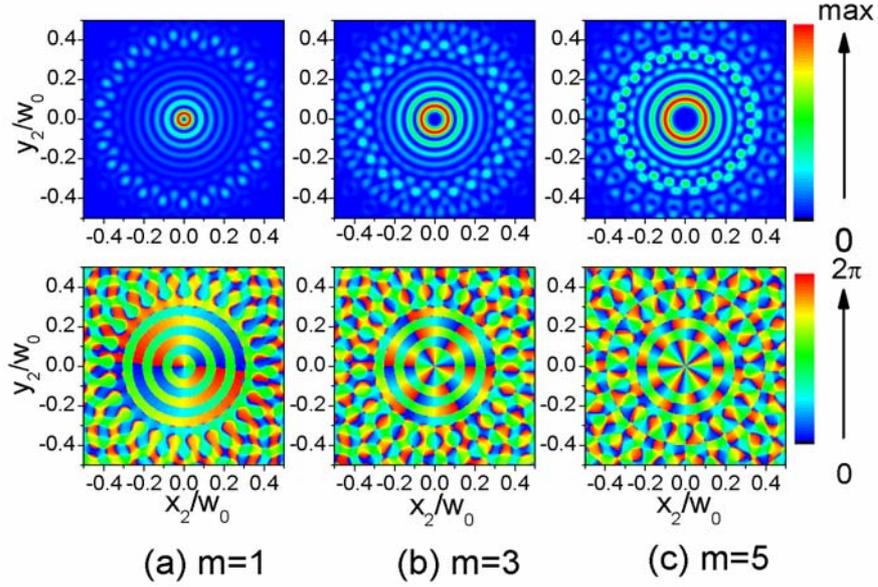

FIG. 8. The intensity (upper) and phase (below) distributions for the different radial beam arrays with different topological charges: (a) $m=1$, (b) $m=3$ and (c) $m=5$ at the focusing plane ($z_1=0$), with other parameters $N=25$ and $\rho=16w_0$.

Figure 8 show the effect of the topological charge $m$ on the focusing properties of the radial beam array. It is clear seen that the innerest dark-hollow region becomes larger and larger for the radial beam arrays, having the initial well-ordered phase distributions of the larger topological charge $m$. It indicates that one can control the initial phases on the beamlets of the initial radial beam array to design the desired output optical vortices with different topological charges. Under the fixed number $N$ and other fixed parameters, with the increasing of $m$ the number for the multi-annular hollow vortex beams with the homocentric phase helical structures becomes smaller and smaller.

## 4. Summary

We have proposed a novel method to generate optical vortices by using the coherent-superposition of Gaussian beamlets in a radial symmetrical configuration. The general propagation formulae for the radial coherent beam arrays passing through the first-order linear optical systems are analytically derived in terms of the generalized Huygens-Fresnel diffraction integral. Based on the derived equations, the propagation properties of the coherent radial beam arrays are analyzed in detail. From the intensity and phase structures, it is found that the resulted beams become optical vortices with the rotating effects and phase helical structures. It should be emphasized that although we only consider the propagation of the radial beam arrays composed of the fundamental coherent Gaussian beams, it is easily to generalize this method into other complex cases such as the beam arrays composed of the coherent Laguerre-Gaussian beams (i. e. similar to vortex arrays [26]) and the arrays composed of the coherent truncated Bessel beams [38]. Finally we have to mention the previous work proposed by Pas'ko et al.[39] where the transversal optical vortex is formed by the interference fields of two Gaussian beams.

**Acknowledgments** This work was supported by the National Nature Science Foundation of China (10604047), by Zhejiang Province Scientific Research Foundation (G20630 and G80611) and by the financial support from RGC of HK Government (NSFC 05-06/01).